\documentclass[a4paper,fleqn,usenatbib,useAMS]{mnras}
\usepackage[T1]{fontenc}
\usepackage{ae,aecompl}

\usepackage{graphicx}   
\usepackage{amsmath}    
\usepackage{amssymb}    
\usepackage{breakurl}   

\newcommand{\UBVR}{$U\!BV\!R$}%
\newcommand{\EBV}{$E(B\!-\!V)$}%

\title[Life after eruption -- VI]{Life after eruption -- VI. 
Recovery of the old novae EL Aql, V606 Aql, V908 Oph, V1149 Sgr, V1583 Sgr 
and V3964 Sgr}
\author[C. Tappert et al.]%
{%
C. Tappert,$^{1}$\thanks{E-mail: claus.tappert@uv.cl},
D. Barria$^{2,3}$,
I. Fuentes Morales$^{1}$,
N. Vogt$^{1}$ ,
A. Ederoclite$^{4}$
and
\newauthor
L. Schmidtobreick$^{5}$\\
$^{1}$Instituto de F\'{\i}sica y Astronom\'{\i}a, Universidad de 
Valpara\'{\i}so, Avda. Gran Breta\~na 1111, 2360102 Valpara\'{\i}so, Chile\\
$^{2}$Departamento de Astronom\'ia, Universidad de Concepci\'on, Casilla 160, 
Concepci\'on, Chile\\
$^{3}$Instituto de Astronom\'ia, Universidad Cat\'olica del Norte, 1270709
Antofagasta, Chile\\
$^{4}$Centro de Estudios de F\'{\i}sica del Cosmos de Arag\'on, Plaza San 
Juan 1, Planta 2, Teruel, E44001, Spain\\
$^{5}$European Southern Observatory, Alonso de Cordova 3107, 7630355 Santiago, 
Chile\\
}

\date{Accepted XXX. Received YYY; in original form ZZZ}

\pubyear{2016}

\begin{document}
\label{firstpage}
\pagerange{\pageref{firstpage}--\pageref{lastpage}}
\maketitle

\begin{abstract}
We report on the recovery of the six old novae EL Aql, V606 Aql, V908 Oph, 
V1149 Sgr, V1583 Sgr and V3964 Sgr, using photometric and spectroscopic data.
Analysing several properties, we find that EL Aql is a good candidate for
an intermediate polar. Furthermore, the system inclination of EL Aql, V606 Aql,
V1583 Sgr and V3964 Sgr appears to be sufficiently high to suggest them as good
targets for time series observations. We also eliminate some previously 
suggested candidates for the post-novae V1301 Aql and V1151 Sgr.
\end{abstract}


\section{Introduction}
\defcitealias{tappertetal12-1}{Paper I}%
\defcitealias{tappertetal13-1}{Paper II}%
\defcitealias{tappertetal13-2}{Paper III}%
\defcitealias{tappertetal14-1}{Paper IV}%
\defcitealias{tappertetal15-2}{Paper V}%

A classical nova eruption plays an important role in the evolution of
cataclysmic variable stars 
\citep[CVs; see the books by][]{warner95-1,hellier01-1}. 
It represents the primary mechanism for
mass-loss in these systems, preventing (most of) them to end up as supernovae 
Ia. Irradiation of the donor star by the eruption-heated white dwarf is
furthermore suspected to cause an increase in the mass-transfer rate, which
would have a strong effect on the evolutionary time-scale 
\citep*{prialnik+shara86-1,sharaetal86-1,kovetzetal88-1,pattersonetal13-2}.
Still, the connection between the nova eruption and the physical parameters
of CVs is not very well known, mainly due to the fact that a large fraction
of the post-nova systems remain unidentified \citep{tappertetal12-1}.

We are carrying out a systematic observing campaign in order to recover and 
study classical novae that erupted at least 30 yr ago (earlier than 1986). 
This implies that sufficient time has passed since the nova eruption for the
optical signal to become dominated by the underlying CV instead of 
representing mainly the ejected material. The targets presented here were 
selected from \citet{downesetal05-1} and refer to fields with hitherto 
ambiguous identification of the post-nova (EL Aql, V606 Aql, V908 Oph, V1149 
Sgr and V1583 Sgr). They were investigated by means of {\UBVR} photometry to 
identify potential candidates for the post-nova, based on their colours. The 
selected candidates were subsequently observed spectroscopically to confirm 
the post-nova via their spectroscopic characteristics (mainly the presence
of characteristic emission lines). In addition, we also obtained spectrograms
of candidates for the post-novae V1301 Aql, V1151 Sgr and V3964 Sgr that were 
included in \cite{tappertetal15-2}. Other collections of post-nova spectra
can be found in \citet*{ringwaldetal96-3}, \citet{tomovetal15-1} and parts
of our own series \citep[][the latter hereafter is referred to as Paper V]%
{tappertetal12-1,tappertetal14-1,tappertetal15-2}.

\section{Observations and reduction}
\label{obs_sec}


\begin{table*}
\begin{minipage}{2.0\columnwidth}
\caption[]{Log of observations.}
\label{obslog_tab}
\setlength{\tabcolsep}{0.15cm}
\begin{tabular}{@{}llllllll}
\hline\noalign{\smallskip}
Object & RA (2000.0) & Dec.~(2000.0) & rms (arcsec) & Date & Filter/grating 
& $t_\mathrm{exp}$ (s) & mag \\
\hline\noalign{\smallskip}
EL Aql    & 18:56:01.87 & $-$03:19:18.8 & 0.21 & 2012-07-18 & $U$/$B$/$V$/$R$
& 1576/420/160/120 & 20.9V \\
          &             &               &      & 2015-04-27 & B600 (1.5 arcsec)
& 4760             & 19.9R \\
V606 Aql  & 19:20:24.29 & $-$00:08:07.8 & 0.30 & 2012-06-24 & $U$/$B$/$V$/$R$
& 1576/420/160/120 & 20.4V \\
          &             &               &      & 2015-06-15 & B600 (1.5 arcsec)
& 3600             & 20.1R \\
V908 Oph  & 17:28:04.58 & $-$27:43:04.4 & 0.33 & 2014-04-03 & $U$/$B$/$V$/$R$
& 1655/495/200/170 & 20.5V \\
          &             &               &      & 2015-04-28 & B600 (1.5 arcsec)
& 3600             & 18.5R \\
V1149 Sgr & 18:18:29.85 & $-$28:17:24.9 & 0.21 & 2012-08-15 & $U$/$B$/$V$/$R$
& 1576/420/160/120 & 18.4V \\
          &             &               &      & 2015-06-06 & B600 (1.5 arcsec)
& 1350             & 18.1R \\
V1583 Sgr & 18:15:26.53 & $-$23:23:17.5 & 0.27 & 2012-07-24 & $U$/$B$/$V$/$R$
& 1576/420/160/120 & 20.5V \\
          &             &               &      & 2015-04-21 & B600 (1.5 arcsec)
& 3600             & 19.5R \\
V3964 Sgr & 17:49:42.42 & $-$17:23:34.8 & 0.17 & 2015-04-26 & B600 (1.5 arcsec)
& 3600             & 19.0R \\
\hline\noalign{\smallskip}
\end{tabular}
\end{minipage}
\end{table*}

The {\UBVR} photometric data for the objects with ambiguous identification
were taken in two service mode runs in 2012 (EL Aql, V606 Aql, V1149 Sgr and
V1583 Sgr) and in 2014 (V908 Oph) at the Antu 8\,m telescope that is part of
the Very Large Telescope of the European Southern Observatory (ESO), 
at Cerro Paranal, Chile, using the FOcal Reducer/low dispersion Spectrograph 
\citep[FORS2;][]{appenzelleretal98-3} system with the high-throughput 
broad-band filters u\_High, b\_High, v\_High and R\_Special. Series of
four and three frames per filter were obtained in the 2012 and the 2014
observations, respectively. In service mode, FORS2 usually operates with
a mosaic of two 2k$\times$4k MIT CCDs. For the data obtained on 2012 August
18, exceptionally, the blue-sensitive E2V CCDs were used. 
The field of view of FORS2 is 6.8$\times$6.8 arcmin$^2$. 

After combining the two mosaic CCD frames into one with the {\sc fsmosaic} 
routine from the FORS Instrument Mask Simulator package,
the further reduction was performed using the {\sc ccdred} package
of {\sc iraf}. The bias-subtracted and flat-fielded data were then corrected
for the telescope offsets between them and averaged to a single frame per
filter. A 3$\sigma$ clipping algorithm was applied to exclude bad pixels and
cosmic rays. 

Subsequent photometry was performed using {\sc iraf}'s {\sc daophot} package 
and the stand-alone {\sc daomatch} and {\sc daomaster} routines 
\citep{stetson92-1}. The aperture radius for the photometry was chosen to
be slightly smaller than the full width at half-maximum (FWHM) of the 
point spread function (PSF). The conversion to calibrated magnitudes was
obtained using observations of standard fields 
\citep{landolt83-1,landolt92-3,stetson00-2}. The $U$ passband is not a part of
ESO's standard calibration plan, and in most nights, the standard field 
observations in this filter covered only a very limited range of airmasses
that did not allow us to derive a value for the extinction from them. Thus,
a standard value for the extinction was used that was taken from the
La Silla observatory 
website\footnote{\url{http://www.eso.org/sci/facilities/lasilla/instruments/efosc/inst/zp/.html}}.
While this observatory is located
roughly 480 km south of Cerro Paranal, comparison of the $BV\!R$ passbands 
and also the $U$ data for the few nights, where a sufficiently large airmass 
range was covered, showed identical values within the uncertainties. Similarly, 
for some nights, also the $BV\!R$ standard data proved insufficient. For 
those filters, median values for a 60 d range are available at the 
observatory's Health Check 
Monitor\footnote{\url{http://www.eso.org/observing/dfo/quality/ALL/daily_qc1.html}}.

The spectroscopic data were obtained using the Gemini Multi-Object 
Spectrograph \citep[GMOS;][]{hooketal04-1} mounted at the Gemini-South
8\,m telescope situated on Cerro Pach\'on, Chile. Three spectra were taken
per object, in most cases subsequently. Since 2014, GMOS is
equipped with a row of three 2048$\times$4176 Hamamatsu CCDs. Grating B600 was
employed with a 1.5 arcsec slit and a two and four pixels binning in spatial
and dispersion direction, respectively. This combination yielded a typical 
wavelength range of 4100 -- 7100 {\AA} and a resolving power of $\sim$800. 
The grating was operated at a central wavelength of 5600 {\AA} to put the two 
0.915 mm wide gaps between the chips at regions of comparatively low scientific 
interest. The typically affected wavelength ranges are 5030 -- 5080 {\AA} and, 
in principle, 6100 -- 6050 {\AA}. However, all observations suffered from
a bright large-scale distortion in spatial direction that severely affected
the signal at a much broader wavelength range of roughly 5900 -- 6200 {\AA}
and could not be properly corrected for. 

The data were reduced using the {\sc gemini -- gmos} extension for {\sc iraf} 
as contained in the {\sc ureka} software 
package\footnote{\url{http://ssb.stsci.edu/ureka/}}. After
subtraction of an averaged bias frame and subsequent division by a flat-field
that was normalized by fitting a high-order cubic spline, a two-dimensional
wavelength correction was applied that was obtained from spectra taken with
a CuAr comparison lamp. After this transformation, the individual spectra were
combined employing a 2$\sigma$ clipping algorithm to correct for cosmic rays.
The sky was subtracted by fitting a Chebychev function of typically third 
order in
spatial direction. Finally, the spectra were extracted and corrected for
the instrumental response function by comparison with a spectrum of the
standard star LTT 3864 taken on 2015 February 25. Since all calibrations
are based on this single observation of a standard, we emphasize that,
while the spectral energy distribution (SED) of our objects should be
sufficiently well reproduced, the zero-point is not well determined, and thus 
the data cannot be regarded as flux calibrated. Above steps were performed
both for the combined and the individual spectra to examine them for
variability.

The details of the observations of the confirmed post-novae are summarized 
in Table \ref{obslog_tab}. The coordinates in columns two and three were 
determined by calculating an astrometric correction on the photometric $R$ 
FORS2 images or the $r^\prime$
GMOS acquisition frames using Starlink's 
{\sc gaia}\footnote{\url{http://astro.dur.ac.uk/~pdraper/gaia/gaia.html}} tool 
(version 4.4.6) with the US Naval Observatory CCD Astrograph Catalogue 
version 4 \citep{zachariasetal13-1}. The associated root mean square
(rms) is given in column four. The fifth column states the date of the
observations referring to the start of the night, while the sixth column
gives details on the instrumental setup, i.e.~filter or grating and slit,
and column seven contains the total exposure time per object. Finally,
the brightness of the object at the time of the observations is presented
in column eight, with the corresponding passband indicated by a letter.
For the spectroscopic data, the magnitudes were determined via differential
photometry on the acquisition frames with respect to five comparison stars
in the vicinity of the target. These values were calibrated with respect to
previous photometric data, where available, or with the
Guide Star Catalogue, version 2.2 \citep{laskeretal08-1}.

\section{Results}
\label{results_sec}

\begin{table}
\caption[]{Results of the {\UBVR} photometry.}
\label{ubvr_tab}
\begin{tabular}{@{}lllll}
\hline\noalign{\smallskip}
Object    & $V$       & $U\!-\!B$   & $B\!-\!V$ & $V\!-\!R$ \\
\hline\noalign{\smallskip}
EL Aql    & 20.88(03) & $-$0.10(15) & 0.62(08)  & 0.70(08)  \\
V606 Aql  & 20.41(03) & $-$0.28(15) & 0.40(07)  & 0.44(07)  \\
V908 Oph  & 20.54(11) & $-$0.10(15) & 1.02(15)  & 1.23(17)  \\
V1149 Sgr & 18.36(05) & $-$1.44(15) & 0.30(12)  & 0.30(07) \\
V1583 Sgr & 20.54(04) & $-$0.16(18) & 0.96(09)  & 0.83(06) \\
\hline\noalign{\smallskip}
\end{tabular}
\end{table}

\begin{figure}
\includegraphics[width=\columnwidth]{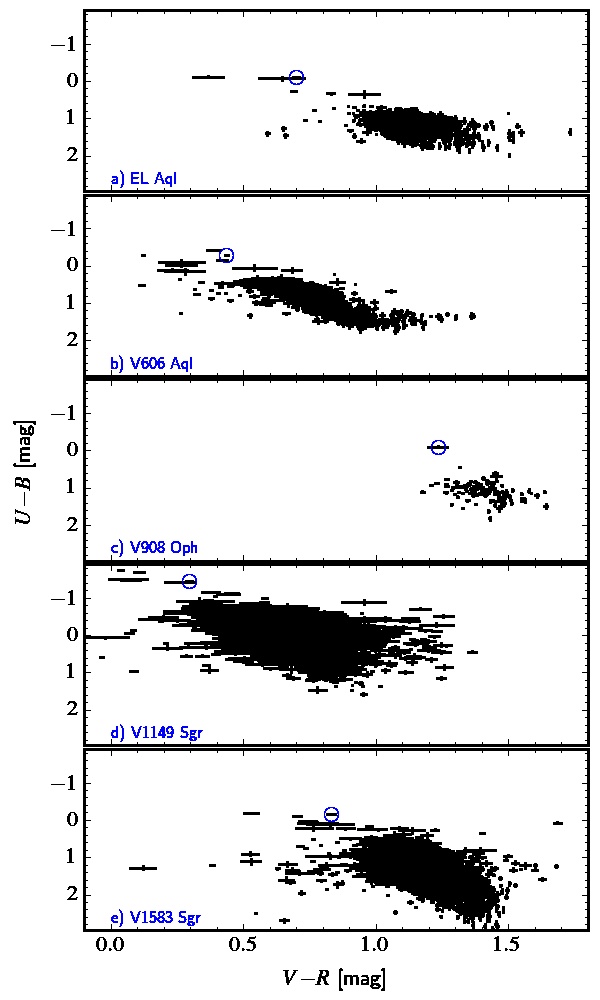}
\caption[]{
Colour--colour diagrams for the five post-nova fields with {\UBVR} 
data. In each field, the post-nova is indicated by a circle.}
\label{ubvr_fig}
\end{figure}

\begin{figure*}
\includegraphics[width=2.0\columnwidth]{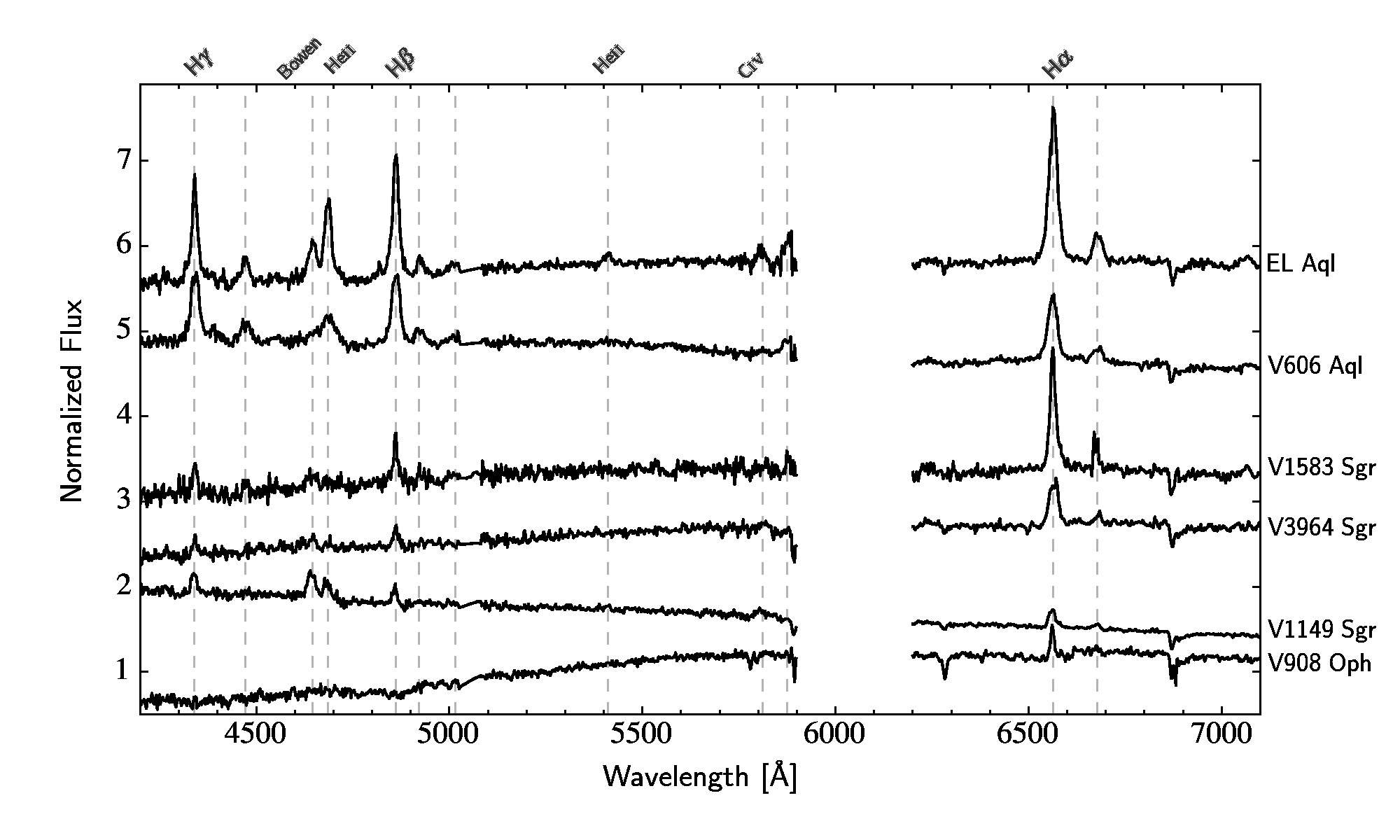}
\caption[]{Spectra of the six post-novae. The data have been divided by
their mean value and displaced vertically for visibility.
Vertical dashed lines mark the principal emission features. Unlabelled lines 
indicate the He\,{\sc i} emission series.
}
\label{allspec_fig}
\end{figure*}

\begin{table*}
\begin{minipage}{16cm}
\caption[]{Equivalent widths in angstroms of the principal emission lines.
Values in square brackets refer to components within absorption troughs.
}
\label{eqw_tab}
\begin{tabular}{@{}llllllllllll}
\hline\noalign{\smallskip}
Object    &  \multicolumn{3}{c}{Balmer} 
& \multicolumn{4}{c}{He\,{\sc i}} & Bowen & \multicolumn{2}{c}{He\,{\sc ii}} 
& C\,{\sc iv}\\
          & 4340     & 4861      & 6563            
& 4922 & 5016 & 6678 & 7065                        
& 4645 & 4686                   & 5412   & 5812 \\ 
\hline\noalign{\smallskip}
EL Aql    & 26(1)    & 35(1)     & 47(1)           
& 4(1) & 3(1) & 8(1) & 5(1)                        
& \multicolumn{2}{c}{45(2)$^1$} & 3(1)   & 4(1) \\ 
V606 Aql  & 18(1)    & 21(1)     & 26(1)           
& 4(1) & 2(1) & 4(1) & 2(1)                        
& \multicolumn{2}{c}{16(1)$^1$} & --     & --   \\ 
V908 Oph  & 1(1)$^2$ & 1(1)$^2$  & 3(1)            
& --   & --   & --   & --                          
& --   & --                     & --     & --   \\ 
V1149 Sgr & 3(1)     & 2(1)      & 6(1)            
& --   & --   & 1(1) & --                          
& 5(1) & 3(1)                   & 1(1)   & 3(1) \\ 
V1583 Sgr & 6(1)     & 9(1)      & 26(5)           
& 3(1) & --   & 4(1) & 2(1)                        
& 5(1) & 2(1)                   & --     & --   \\ 
V3964 Sgr & 3(1)     & 5(1)      & 12(1)           
& 1(1) & --   & 2(1) & --                          
& 3(1) & 2(1)                   & --     & --   \\ 
\hline\noalign{\smallskip}
\end{tabular}
\\
$^1$ Bowen/\mbox{He\,\sc ii} blend.\\
$^2$ Emission cores in absorption troughs.
\end{minipage}
\end{table*}

Colour--colour diagrams for the fields of reported novae with ambiguous
identification for the post-nova are presented in Fig.~\ref{ubvr_fig}. 
We have excluded objects with photometric errors $>$0.1 mag or
$<$0.005 mag to minimize contamination by very faint sources and artefacts,
respectively. Candidates for subsequent spectroscopy were selected based on 
their position
in the colour--colour diagram with respect to the other stars in the field. 
An additional selection criterion was the distance to the reported coordinates
of the nova, taking into account their detection history. For example, the
search area for novae that were identified on photographic plates was limited
to the central 1 arcmin$^2$. Subsequent long-slit spectroscopic data 
were used to confirm the post-nova. The spectra of the confirmed novae are
presented in Fig.~\ref{allspec_fig}, their photometric data are summarized in 
Table \ref{ubvr_tab}. 

We briefly describe the methods used to analyse the spectroscopic properties
that apply to all targets. The emission line strength is characterized via
the equivalent width $W_\lambda$, which is calculated with respect to three 
manually selected wavelength ranges. Two of them are used to define the 
continuum and the noise on either side of the line, while the third one 
contains the line itself. We then integrate over the range of the latter
divided by a linear fit to the continuum. The uncertainty of the resulting
$W_\lambda$ is estimated via a Monte Carlo simulation by repeating the
measurement a thousand times to a variation of the data with respect to
the addition of a random value within the noise interval. The resulting
equivalent widths are presented in Table \ref{eqw_tab}.

We make an attempt at recovering the intrinsic SED by correcting for
the interstellar extinction in the line of sight. We use the interface on
NASA's Infrared Science Archive (IRSA) web 
pages\footnote{\url{http://irsa.ipac.caltech.edu/applications/DUST/}} to 
obtain values {\EBV} from \citet{schlafly+finkbeiner11-1}. These serve
as input for the dereddening routine implemented in {\sc iraf} and based
on the relations from \citet*{cardellietal89-1}. A standard value
for the ratio of the total to the selective extinction 
$R(V) = A(V)/E(B\!-\!V) = 3.1$ is employed. In order to characterize the
corrected continuum slope, we fit the wavelength range between 5000 and
7000 {\AA} with a power law $F \propto \lambda^{-\alpha}$. For the intrinsic
SED, the exponent $\alpha$ then depends on the mass-transfer rate and
the system inclination. Larger values for $\alpha$ correspond to higher
mass-transfer rates (hotter and thus bluer discs) and lower inclinations
(a larger fraction of the hotter inner disc is visible). However, there
are large uncertainties involved in the reddening correction, because these
are large-scale extinction maps with a resolution of 2$\times$2 deg and local
fluctuations may be present. Furthermore, the exact depth position of the
nova with respect to the absorbing material is also unknown. These 
uncertainties propagate to the derived $\alpha$, which should thus be taken 
only as an approximate indicator of the intrinsic SED.

\subsection{EL Aquilae = Nova Aql 1927}
\label{elaql_sec}

\begin{figure*}
\includegraphics[width=2.\columnwidth]{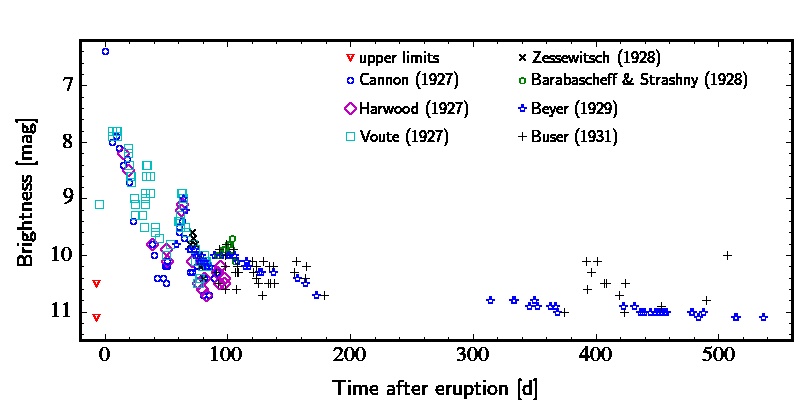}
\begin{minipage}{2.0\columnwidth}
\caption[]{Decline light curve of EL Aql.}
\label{elaqllc_fig}
\end{minipage}
\end{figure*}

\begin{figure}
\includegraphics[width=\columnwidth]{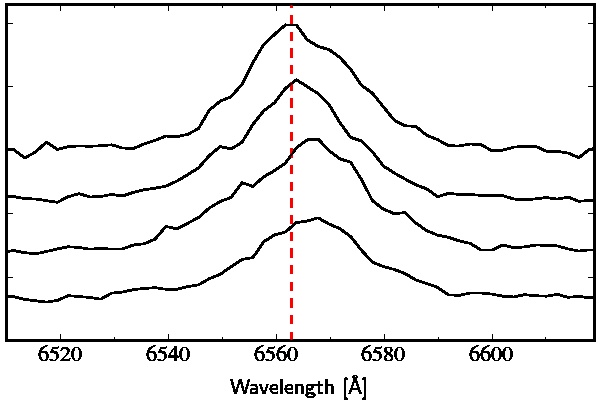}
\caption[]{H\,$\alpha$ line profiles for the four individual spectra of EL Aql
in sequence from bottom to top. The vertical line indicates the rest wavelength
of H\,$\alpha$.}
\label{elaqllprof_fig}
\end{figure}

The nova was discovered by \citet{wolf27-3} as a star of ninth magnitude
on photographic plates taken on 1927 July 30 and 31. Examination of
pre-discovery Harvard plates by \citet{cannon27-9} revealed the object having
reached its peak brightness before that date. The brightest recorded
value is 6.4 mag on 1927 June 15, which is thus taken as maximum
brightness, although observational gaps before and after that date leave
room for an even brighter peak value. The decline of the nova is comparatively
well covered, especially in the first 200 d after eruption. Description of
the spectral evolution during the decline is found in a number of publications.
The most detailed data can be found in \citet{wyse40-2}, while 
\citet{schajn+nikonoff28-1} present the only drawing of the spectrum.

In spite of the good photometric coverage, there is no complete representation
of the long-term light curve in the literature. The plot in \citet{wyse40-2}
is limited to the first 100 d, and \citet{beyer29-3} does not include the
Harvard data. In Fig.\,\ref{elaqllc_fig}, we collect the observations from
\citet{cannon27-9}, \citet{harwood27-5}, \citet{voute27-11}, 
\citet{zessewitsch28-2}, \citet{barabascheff+strashny28-1}, 
\citet{beyer29-3} and \citet{buser31-2}. The dates and brightness values
from \citet{voute27-11} were extracted from his plot with a precision of
0.25 d and 0.1 mag, respectively. All other data were available in tabulated
form. For data where only the start of night was given, {\sc UT} = 0 h was 
assumed
for the observation. We omitted from the plot the first four data points from 
\citet{buser31-2}, where he describes the target as `blurred', because they
differ drastically from those of other authors. The upper limits in the plot
are from \citet{cannon27-9} and \citet{voute27-11}. For reasons of
visualization, we omit another upper limit ($>$16 mag) from 1927 May 28
\citep{cannon27-9}.

After the initial decline, the nova experienced a series of minor
rebrightenings and then settled on a prolonged plateau phase. There also
appear to be further fluctuations around 400 and 500 d after eruption.
These are based on the \citet{buser31-2} data only, but since in the 
overlapping regions his magnitudes usually agree within 0.2 mag to those 
of other authors, there is no reason to doubt the reality of these 
rebrightenings.
To our knowledge, there are no further data on the nova after the seasonal gap
between 540 and 720 d after eruption. Assuming that this is not due to the lack
of observations, the final decline must have occurred sometime during that
period. \citet{wyse40-2}
observed the field again on 1940 June 5 and reported a brightness of 19 mag
for the nova, but, as we argue below, this is a likely misidentification.

The decline light curve is classified in \citet{duerbeck81-1} as `Ba' that
includes novae with `standstills or other minor irregular fluctuations
during decline'. Although EL Aql is not included in the \citet*{stropeetal10-1} 
catalogue, its light curve likely corresponds to the `P' (plateau) 
class, and in particular, shows some similarity to HS Sge, especially with
respect to the rebrightenings. In EL Aql, the major peak around day 63 makes
it difficult to define the end of the initial decline and the start of the
plateau phase. If we select the end of this event as the corresponding point,
we find the following approximate values for the light-curve parameters as 
defined by \citet{stropeetal10-1}: the time of the start of the plateau phase, 
and thus the duration of the initial decline, 
$\Delta T_\mathrm{plat} \sim$87 d, the duration of the plateau phase
$D_\mathrm{plat} \sim$450--630 d, the brightness difference between the
start of the plateau phase and the peak magnitude 
$\Delta V_\mathrm{plat} \sim$3.9 mag, and the slope of the decline during
the plateau phase to $\sim -0.3$. These are typical values for class P
novae and justify the inclusion of EL Aql into this category. 
\citet{stropeetal10-1} find that this class contains a large fraction 
(60--90\%) of the recurrent novae and suggest that some of the other systems
in this class are yet unrecognized recurrent novae. The remaining members
of the P class novae appear to be magnetic systems. 

A candidate for the post-nova was selected from our {\UBVR} photometric
data on the basis of its colours (Fig.~\ref{ubvr_fig}, Table \ref{ubvr_tab})
and its proximity (3 arcsec) to the coordinates reported in 
\citet{downesetal05-1}. The spectrum of this object presents emission lines 
typical for a CV and we conclude that it is indeed the post-nova. Its 
brightness is $V = 20.9$ mag and thus considerably fainter than the 19th 
magnitude star observed by \citet{wyse40-2} in 1940.
\citet{szkody94-2} gives a value of $V = 18.5$ mag for EL Aql for observations
made in 1989. It is, in principle, possible that both observations above
caught the post-nova in a bright state (or our data were taken in a low
state). However, we consider this unlikely, because CVs in bright states are 
also expected to have bluer colours than in low states, and the object
observed by \citet{szkody94-2} is significantly redder in the visual
range ($B-V = 1.34$ mag compared to our 0.62 mag, see Table \ref{ubvr_tab}).
We thus conclude that those observations did not concern the post-nova,
but a different object. Similarly, \citet{harrison92-1} gives infrared data
$K = 12.55$ mag, $J-H = 0.95$ mag, $H-K = 0.17$ mag for EL Aql. This led
\citet{pagnotta+schaefer14-1} to suspect the nova to contain a giant
secondary star. However, the spectral appearance of the post-nova 
(Fig.~\ref{allspec_fig}) and its position in the colour--colour diagram
does not support this interpretation, and it thus again appears more likely
that \citet{harrison92-1} observed a different object. We furthermore note
that the resulting eruption amplitude (defined as the difference between
the brightness of the post-nova and the maximum brightness) of 14.5 mag 
basically excludes EL Aql as a recurrent nova candidate.

EL Aql presents a rich emission line spectrum (Fig.~\ref{allspec_fig}, Table
\ref{eqw_tab}). The Balmer lines are exceptionally strong for an old nova,
which might be indicative of a low mass-transfer rate \citep{patterson84-1}.
The strength of He\,{\sc ii} $\lambda$4686 and the presence of He\,{\sc ii} 
$\lambda$5412 points to a magnetic system, but since the former does not
quite match the strength of H$\beta$, EL Aql is more likely to be an 
intermediate polar rather than a discless system. 
The continuum has a slightly red slope,
but as can be seen from the comparison of the colour--colour diagrams
(Fig.~\ref{ubvr_fig}), the field of EL Aql is significantly affected by
interstellar reddening. From the IRSA website, we find {\EBV} = 1.11(08) mag,
and when corrected for that value, an inverse power law with an exponent 
$\alpha = 3.12(02)$ provides an excellent fit to the continuum slope.

The spectroscopic data on EL Aql consist of four individual spectra. In
Fig.~\ref{elaqllprof_fig}, we examine the variation of the H\,$\alpha$ line
profile over the course ($\sim$1 h) of our observations. We find that the
line as a whole shows a significant Doppler shift. Estimating the shift
in radial velocity units by fitting a single Gaussian function to the
line profiles yields a variation of $\Delta v_\mathrm{r}$\,$\sim$\,100 
km s$^{-1}$ over
the observed time range. Furthermore, also the shape of the line profile 
appears to be variable, pointing to the presence of an additional emission 
source in the system, such as from an irradiated secondary star or a bright 
spot \citep[e.g.][]{tappert+hanuschik01-1}.

From the gathered evidence, we conclude that EL Aql is a likely intermediate
polar seen at a moderately high inclination. In spite of its faintness,
its orbital period might therefore be accessible with time series photometry.

\subsection{V606 Aquilae = Nova Aql 1899}
\label{v606aql_sec}

\begin{figure}
\includegraphics[width=\columnwidth]{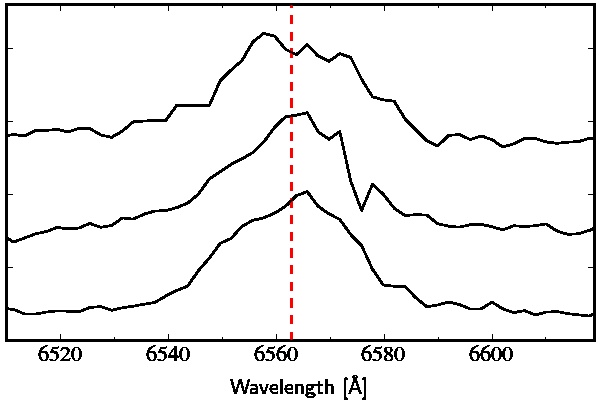}
\caption[]{H\,$\alpha$ line profiles for the three individual spectra of V606 Aql
in sequence from bottom to top. The vertical line indicates the rest wavelength
of H\,$\alpha$.}
\label{v606aqllprof_fig}
\end{figure}

This is a little observed nova that was discovered by Fleming as an object
of seventh magnitude on a photographic plate from 1899 April 21, as reported by 
\citet{pickering00-35}. The brightness of the nova on this plate was later 
more precisely determined to 6.75 mag by \citet{leavitt33-1}. There is a
$\sim$200 d gap of observations prior to the discovery, and it is thus
possible that the real maximum has been missed. \citet{duerbeck87-1}
estimates the peak brightness to 5.5 mag. The observed decline light curve as 
shown in \citet{leavitt33-1} covers $\sim$160 d, and suggests a
classification as type `P' \citep{stropeetal10-1}. In 1991, 
\citet{ringwaldetal96-3} obtained spectroscopy of the suspected post-nova,
but found a G7--K4 type spectrum for a $V$ = 16.3 mag object.

The south-western component of an unresolved visual binary that was marked in 
\citet{downesetal05-1} as the possible post-nova showed promising colours
(Fig.~\ref{ubvr_fig}, Table \ref{ubvr_tab}). Subsequent spectroscopy of
the $V$ = 20.4 mag object confirmed the nova (Fig.~\ref{allspec_fig}). The 
spectrum presents comparatively strong Balmer emission lines, although not as
strong as in EL Aql. The high excitation lines are even considerably weaker,
and we do not detect He\,{\sc ii} $\lambda$5412. As discussed in Section
\ref{elaql_sec}, most members of the `P'-class appear to be recurrent or
magnetic novae. Even if we use the actually observed brightness maximum
to calculate the eruption amplitude, the result of $\Delta m = 13.6$ mag 
appears much too large for a recurrent nova \citep{pagnotta+schaefer14-1}.
On the other hand, the spectroscopic properties do not present particularly
strong evidence for a magnetic nature either. Apparently, V606 Aql is an
unusual member of the P class.

The spectrum presents a slightly blue continuum slope, and the colour--colour
diagram shows that the field is only moderately affected by interstellar
extinction. After dereddening with a value {\EBV} = 0.35(03) mag, it
became apparent that the continuum was distorted by a number of large
`bumps'. After examining the three individual spectra, we concluded that the
bumpy structure is not intrinsic but an artefact. However, the presence of
the bumps rendered large parts of the continuum unusable for fitting. We
thus chose a number of small regions of continuum between the blue
emission lines as well as a couple of regions around 6000 {\AA} to determine
an inverse power-law exponent of $\alpha =$ 2.02(02).

When examining the H\,$\alpha$ line profile of the individual spectra, the
changes are not as pronounced as in EL Aql, and the Dopper shift of the 
whole profile is not as large (Fig.~\ref{v606aqllprof_fig}). Still, the shape 
of the line profile is clearly variable, and the width of the emission lines 
is basically identical to those in EL Aql. We thus tentatively conclude that 
also V606 Aql is likely seen at a moderately high inclination, which makes it 
a potential target for follow-up time series observations.

\subsection{V908 Ophiuchi = Nova Oph 1954}
\label{v908oph_sec}

\begin{figure}
\includegraphics[width=\columnwidth]{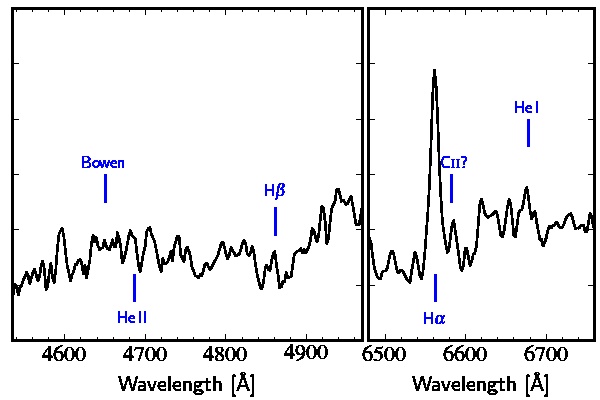}
\caption[]{Close-up on the line profiles in two selected wavelength ranges
in the spectrum of V908 Oph.}
\label{v908ophlprof_fig}
\end{figure}

Very few data are available for V908 Oph. Its discovery was reported by
\citet{blanco54-1} who detected O\,{\sc i} emission in a near-infrared
spectrum taken on 1954, July 2. He estimated the brightness in this
spectral range to 9 mag, but suspected that the nova was observed about a
week after its maximum brightness. The only other observations consist of
two objective prism spectra taken on 1954 July 27 by \citet{seyfert+yoss54-1}.
Apart from strong H\,$\alpha$ emission, they could not distinguish more
detailed spectral features due to the object having declined to a blue
magnitude `probably less than 14'. If the object was indeed observed 
several days after having reached its maximum brightness, and accounting
for the colour differences between the two observations and for the
strong interstellar reddening, the time $t_3$ to have declined
by 3 mag below maximum brightness is likely to be in the order of 30--40 d, 
which would classify the object as a moderately fast nova.

The coordinates of the nova had only been recorded with a precision in
the order of arcmin, but fortunately, a candidate for the post-nova can be 
easily identified in the colour--colour diagram (Fig.~\ref{ubvr_fig}). Its 
spectrum presents a few weak hydrogen emission lines, with those bluewards
of H\,$\alpha$ being embedded in absorption troughs, and a red continuum slope 
(Fig.~\ref{allspec_fig}). From the {\UBVR} photometry, it is clear that the 
field is 
strongly affected by interstellar reddening, and the respective catalogue
yields an accordingly high {\EBV} = 1.26(02) mag. Once corrected for this
value, the SED corresponds to a steep blue power law with a negative
exponent $\alpha = 3.38(02)$. Thus, the spectral characteristics are
consistent with a high mass-transfer system. In Fig.~\ref{v908ophlprof_fig},
we show the soectral regions around H\,$\beta$ and H\,$\alpha$. in more detail.
We note that high excitation emission lines like the Bowen blend and 
He\,{\sc ii} are absent or, at least, very weak. This is in stark contrast to
other likely high mass-transfer post-novae like GR Sgr 
\citepalias{tappertetal15-2}.

Towards the red side of H\,$\alpha$ emission there appears to be another
peak that could be identified with the C\,{\sc ii} $\lambda\lambda$6578,6583
doublet. This line has been previously detected in a few old novae with
sufficiently narrow H\,$\alpha$ emission lines \citep{tappertetal13-2}, and
the latter also applies to V908 Oph. However, the signal-to-noise ratio
(S/N) in that part of
the spectrum is low, and there are a number of similar large features close-by,
which we were unable to identify and thus suspect to be artefacts. In
consequence, the detection of C\,{\sc ii} in this object remains ambiguous.

We finally remark that we did not detect any significant positional
variation of the H\,$\alpha$ line in the three individual spectra. Together
with the narrowness of that line, this suggests that the system is seen at low 
inclination, and that it will be difficult to determine its orbital period.

\subsection{V1149 Sagitarii = Nova Sgr 1945}
\label{v1149sgr_sec}

\begin{figure}
\includegraphics[width=\columnwidth]{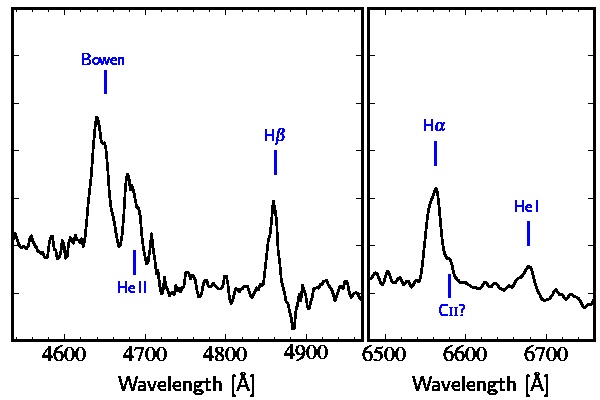}
\caption[]{Close-up on the line profiles in two selected wavelength ranges
in the spectrum of V1149 Sgr.}
\label{v1149sgrlprof_fig}
\end{figure}

The nova was discovered by \citet{mayall49-1} on an objective prism plate at
a photographic brightness of about 9 mag. Later research by \citet{warren65-1}
revealed the maximum brightness to 7.4 mag. The light curve is described as
a `very steep increase in brightness [...] followed by a rapid initial
decrease', reaching a plateau by 1946 May 28. The nova was visible from 
1945 May 16 to 1948 May 30, and afterwards fainted below 14.7 mag. Large gaps
in the observations prevent a more detailed analysis.

From our {\UBVR} photometry (Fig.~\ref{ubvr_fig}), we selected two potential 
candidates for the post-nova based on their colours and coordinates.
Spectroscopy showed one of those to present emission lines and a blue
continuum, and we thus identify this object as the post-nova. It is located
$\sim$11 arcsec south-west of the previously reported position. We note that
these coordinates do not coincide with the objects labelled as V1149 Sgr in 
\citet{saitoetal13-2} and in \citet{mrozetal15-3}. 

The spectrum shows a blue continuum, that becomes somewhat steeper 
($\alpha = 2.41(01)$) when corrected for the catalogued interstellar reddening 
of {\EBV} = 0.39(03) mag, and weak 
emission lines (Fig.~\ref{allspec_fig}). In contrast to V908 Oph (Section 
\ref{v908oph_sec}), V1149 Sgr counts with a strong Bowen blend and He\,{\sc ii} 
$\lambda$4686 emission (Fig.~\ref{v1149sgrlprof_fig}). Since no other 
He\,{\sc ii} lines can be identified, the system is not a strong candidate for
a magnetic CV. The Balmer lines are comparatively narrow, and a bump on the 
red wing of H\,$\alpha$ could tentatively identified with C\,{\sc ii} (see also
Section \ref{v908oph_sec}). The three individual spectra do not show any
significant radial velocity variations, and thus, like V908 Oph, this object
is likely to be seen at a rather low system inclination.

\subsection{V1583 Sagitarii = Nova Sgr 1928}
\label{v1583sgr_sec}

\begin{figure}
\includegraphics[width=\columnwidth]{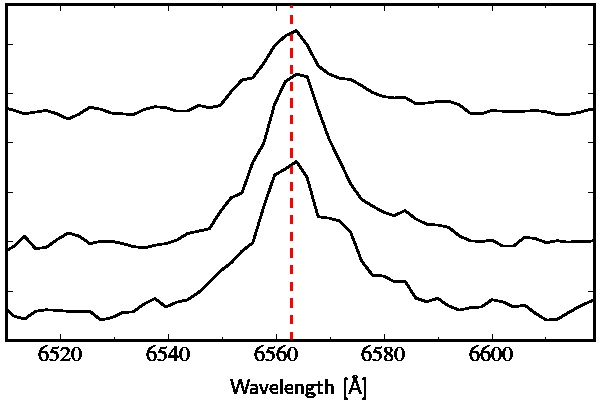}
\caption[]{H\,$\alpha$ line profiles for the three individual spectra of V1583
Sgr in sequence from bottom to top. The vertical line indicates the rest 
wavelength of H\,$\alpha$.}
\label{v1583sgrlprof_fig}
\end{figure}

An examination of archival photographic plates by \citet{dishong+hoffleit55-1}
led to the discovery of this nova almost 30 yr after its eruption. Still,
in contrast to other novae, the maximum brightness of 8.9 mag on 1928 June
24 appears to be well established. The authors report that two days before 
that night, the nova was fainter than 15.6 mag and that it fell again below the
plate detection limit 75 d after maximum brightness. The system is classified
as a moderately fast nova with a decline time $t_3 = 37$ d 
\citep{duerbeck87-1}. 

From its colours (Fig.~\ref{ubvr_fig}, Table \ref{ubvr_tab}) and its position 
only 3.5 arcsec off the reported coordinates, we identified a candidate for 
the nova. Subsequent spectroscopy show a slightly reddish continuum
(Fig.~\ref{allspec_fig}), which is likely the consequence of a comparatively 
strong interstellar reddening ({\EBV} = 1.09(04) mag). Like for most of the 
other systems, the corrected SED presents a blue slope, with 
$\alpha = 3.27(02)$. Superposed are 
moderately strong emission lines (Table \ref{eqw_tab}) of the Balmer and 
He\,{\sc i} 
series. Similar to V908 Oph (Section \ref{v908oph_sec}), the Bowen blend
and He\,{\sc ii} $\lambda$4686 lines are weak and hardly detectable. 

The emission lines are also comparatively narrow, but still the line profile
shows some clear asymmetric structure and marked differences between the 
individual exposures (Fig.~\ref{v1583sgrlprof_fig}). We especially 
notice that the line strength in the third spectrum of that series is
significantly diminished, which indicates that the emission source is being
obscured. A possible explanation for this apparent discrepancy between narrow
lines and an obscuration effect is that the system is seen at high
inclination, but that the Doppler-broadened component of the line profile 
produced by the disc is weak and that the line profile is dominated by a 
narrow component that originates on the secondary star. In any case, the
observed obscuration makes V1583 Sgr an attractive target for time series 
photometry.

\subsection{V3964 Sagittarii = Nova Sgr 1975}
\label{v3964sgr_sec}

\begin{figure}
\includegraphics[width=\columnwidth]{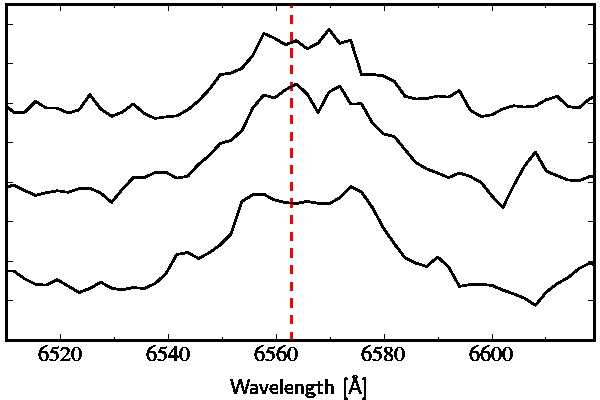}
\caption[]{H\,$\alpha$ line profiles for the three individual spectra of V3964
Sgr in sequence from bottom to top. The vertical line indicates the rest 
wavelength of H\,$\alpha$.}
\label{v3964sgrlprof_fig}
\end{figure}

In \citetalias{tappertetal15-2}, we had observed a candidate for this
nova that turned out to be a K1V star. However, examination of the 2D
spectra showed the possibility of H\,$\alpha$ emission from a different star
in the slit, and a likely source of this emission was identified. Here, 
we present further spectroscopic observations of that target. The spectrum 
in Fig.~\ref{allspec_fig} clearly reveals it as the post-nova system, 
presenting the typical hydrogen and
helium emission lines, as well as a weak Bowen blend. The continuum has
a red slope, but like in the other objects, this is due to interstellar
reddening. Still, even after the correction for {\EBV} = 0.56(01) mag, the 
slope appears significantly
flatter than in the other systems, being represented by an exponent
$\alpha$ = 1.43(03). The emission lines are not particularly strong
(Table \ref{eqw_tab}), but broad, 
and even show, at times, a hint of a double-peaked disc profile
(Fig.~\ref{v3964sgrlprof_fig}). Furthermore, like in V1583 Sgr
(Section \ref{v1583sgr_sec}), the emission line in the third spectrum is
significantly diminished compared to the other two spectra. On the other hand,
a possible Doppler shift of the line is less clear.
Still, the width of the line, the potential obscuration effect and the
flat slope of the corrected continuum suggest that V3964 Sgr is seen at
a comparatively high inclination. Thus, it should be feasible to determine
the orbital period with time series photometry.

\subsection{V1301 Aquilae and V1151 Sagitarii}

Together with V3964 Sgr (Section \ref{v3964sgr_sec}), we had also reported on 
candidates for the novae V1301 Aql and V1151 Sgr that turned out to be 
misidentifications \citepalias{tappertetal15-2}.
For both targets, we had indicated alternative candidates in the immediate
vicinity of the original targets. In the current observing run, we 
incorporated some of these new candidates, namely the objects marked as
`W1' and `W2' for V1301 Aql, and the one marked as `1' in the case of
V1151 Sgr. All objects show rather unremarkable spectra with reddish
continua and Balmer lines in absorption. `1' turned out to be a close
visual binary, but neither component presented any emission lines.
These two post-novae thus remain unidentified. Likely, additional photometric 
data will be needed to select further candidates for spectroscopy.

\section{Discussion}
\label{disc_sec}

\begin{table*}
\begin{minipage}{2.0\columnwidth}
\caption[]{Several properties of the six post-novae. See Section 
\ref{disc_sec} for details.}
\label{prop_tab}
\begin{tabular}{@{}lllllllllllllllll}
\hline\noalign{\smallskip}
Object & $m_\mathrm{max}$$^1$ & $m_\mathrm{min}$$^1$ & $\Delta m$ & $\Delta t$ 
& $t_3$ & {\EBV} & $\alpha$ & $W_\lambda(\mathrm{H}\alpha)$ &
\multicolumn{2}{c}{FWHM({H}$\alpha$)} & $\Delta_\mathrm{pos}$ \\
 & (mag) & (mag) & (mag) & (yr) & (d) & (mag) & & ({\AA}) & ({\AA}) 
& (km s$^{-1}$) & (arcsec) \\
\hline\noalign{\smallskip}
EL Aql    & 6.4$p$    & 20.9$V$ & 14.5    & 88  & 25     & 1.11 & 3.12
& 47 & 27 & 1230 &   3  \\
V606 Aql  & 6.8$p$    & 20.4$V$ & 13.6    & 116 & 65     & 0.35 & 2.02
& 26 & 28 & 1280 &   2  \\
V908 Oph   & $<$9$r$  & 20.5$V$ & $>$11.5 & 61  & $\ge$30 & 1.26 & 3.38
& 3  & 12 & 550  & 234  \\
V1149 Sgr & 7.4$p$    & 18.4$V$ & 11.0    & 70  & $<$270 & 0.39 & 2.41
& 6  & 21 & 960  &  11  \\
V1583 Sgr & 8.9$p$    & 20.5$V$ & 11.6    & 87  & 37     & 1.09 & 3.27
& 26 & 17 & 780  &   4  \\
V3964 Sgr & $<$8.5$p$ & 19.0$R$ & $>$11.5 & 40  & 32     & 0.56 & 1.43
& 12 & 29 & 1320 &   3  \\
\hline\noalign{\smallskip}
\end{tabular}
\\
$^1$ $p$: photographic, $V$: $V$ band, $R$: $R$ band, $r$: near-infrared.\\
\end{minipage}
\end{table*}

In Table \ref{prop_tab} we have collected several distinguished properties
of the here presented post-novae. These are the maximum brightness 
$m_\mathrm{max}$, the brightness of the post-nova $m_\mathrm{min}$ at the
time of our observations and the resulting eruption amplitude defined as
$\Delta m = m_\mathrm{min} - m_\mathrm{max}$. In the next two columns, we give
the time $\Delta t$ that has passed between the nova eruption and our 
spectroscopic data and the time $t_3$ by which the nova had declined to
3 mag below maximum brightness. The parameters in the subsequent two columns 
are associated with the determination of the continuum slope, representing
the catalogued value for the interstellar reddening {\EBV} and the negative
exponent $\alpha$ of the fitted power law (Section \ref{results_sec}). The
next three columns give selected properties of the H\,$\alpha$ emission line:
the equivalent width $W_\lambda(\mathrm{H}\alpha)$ and the FWHM of a
Gaussian fit to the line, the latter both in wavelength and in velocity
units. The final column states the deviation of the position of an object
from the originally recorded coordinates of the nova eruption as given by
\citet{downesetal05-1}.

Since the absolute maximum 
brightness of novae shows a scatter of typically only 2--3 mag 
\citep{dellavalle+livio95-1,kasliwaletal11-2}, but the absolute brightness
of CVs in quiescence is distributed over a range of $\sim$9 mag
\citep[e.g., ][]{patterson84-1}, 
differences in $\Delta m$ will mostly reflect the difference in the brightness
of the post-nova. The latter principally depends on the brightness of the
accretion disc (or, more generally, on the emission from the accretion
process), and thus on the mass-transfer rate $\dot{M}$ and on the system
inclination $i$ \citep{warner86-7}. 
Because $m_\mathrm{max}$ can be assumed to be largely
independent of these two quantities, $\Delta m$ can be taken as an indicator
of the absolute brightness of the post-nova and of the involved parameters.
Apart from the spread in the absolute maximum brightness, additional
uncertainties are introduced by the possibility that the recorded 
$m_\mathrm{max}$ does not correspond to the real maximum brightness, but
that the nova was observed at some later point, and by the different passbands 
used for $m_\mathrm{max}$ and for $m_\mathrm{min}$. While in general,
the colour differences in the optical range for CVs will amount to less than
1 mag \citep[e.g.,][]{szkody94-2}, these differences will be enhanced in
regions with strong interstellar reddening. A thorough analysis of $\Delta m$
will thus have to correct for these effects. However, here we only aim at
a rough classification of our post-novae within the CV subtypes, and
we will do so by also considering other parameters, thus reducing the weight
of above uncertainties.

Further quantities of Table \ref{prop_tab} that are indicative of $\dot{M}$
are $\alpha$ and $W_\lambda(\mathrm{H}\alpha)$. The significance of the former 
and the involved uncertainties have already been presented in Section 
\ref{results_sec}. The indicative power of the latter stems from 
low-$\dot{M}$, optically thin discs producing stronger emission lines than
high-$\dot{M}$, optically thick ones \citep{tylenda81-1,patterson84-1}. 
However, in CVs, typically the accretion disc is not the sole contributor to the 
emission line profile \citep{tappert+hanuschik01-1}. Because post-novae
contain an eruption heated white dwarf, irradiation of the secondary star is 
likely to produce an additional emission component. Furthermore, there is
also the possibility that the ejected material still contributes to the
line profile especially in `younger', i.e.~more recent post-novae. Finally, 
because 
the continuum emission depends on $i$, so does $W_\lambda(\mathrm{H}\alpha)$
\citep{warner87-4}, 
and thus in our interpretation, we will also use the width of the line FWHM,
which is indicative of $i$ due to the larger projected velocity values
$v \sin i$ sampled in the line profile. 

EL Aql and V606 Aql are the novae with the largest eruption amplitudes in
the present sample. Indeed, a comparison with fig.\,7 in 
\citetalias{tappertetal15-2} shows that there are only $\sim$10 novae in our
sample of pre-1986 eruptions with
similar large $\Delta m$. The idea of this parameter being an indicator for
the disc brightness in these two systems fits well with the comparatively
strong emission lines. On the other hand, the lines are also broad, and
thus additionally, the inclination will have contributed to the high value
of $\Delta m$. While, for V606 Aql, also $\alpha$ is in agreement with a
comparatively low-$\dot{M}$ disc, this is not the case for EL Aql. However,
we note that the steepest continuum slopes are found for the systems
with the largest interstellar reddening. This appears suspicious and suggests
a tendency to overcorrect for this effect. 

V908 Oph and V1149 Sgr have the smallest $\Delta m$ and are also
spectroscopically consistent with an optically thick disc and thus high 
$\dot{M}$ in that they present only weak emission lines. In V908 Oph, the 
bluer hydrogen lines are even embedded in absorption troughs. Still, also
here the line width suggests that the inclination is such that it amplifies 
that effect. In conclusion, because of the strong dependence of $\Delta m$
and $W_\lambda$ on $i$, those four post-novae could have very similar 
intrinsic properties with respect to $\dot{M}$ in spite of their different
spectroscopic appearance. 

The picture is somewhat different for the remaining two systems. V1583 Sgr
presents narrow, but comparatively strong emission lines with a variable
line profile and medium $\Delta m$. We interpret this as the combination
of a broad but weak disc line profile and a narrow component that likely
originates in the secondary star. In this way, all parameters would
agree with V1583 Sgr having a medium to high $\Delta m$ and a medium to
high inclination. Assuming that the narrow component is produced by
irradiation, we would also expect that either the white dwarf in this
system is still very hot or that the orbital period is comparatively short.
Because of the `age' of the post-nova ($\Delta t$ = 87 yr), the latter
appears more likely. Finally, V3964 Sgr has broad and weak lines and shows
a possible obscuration effect in one spectrum. This points to a high-$\dot{M}$
system seen at high inclination. The recorded medium $\Delta m$ represents
only a lower limit that, if the speculations by \citet{lundstrom+stenholm76-1,%
lundstrom+stenholm77-1} are correct, could be up to 3 mag larger 
\citepalias[see also][]{tappertetal15-2}. Thus, it would also be in agreement
with a high $i$. Last, not least, the continuum slope appears rather flat
for the suspected high $\dot{M}$, but again might reflect a high system
inclination. 

In \citetalias{tappertetal15-2}, we have argued that there is a high probability
that the CVs detected by us do, in fact, represent the post-novae corresponding
to the respective nova eruption and not different CVs that coincidentally
have similar coordinates. Our first point was that the low space density
of CVs \citep[][and references therein]{pretorius14-1} makes it unlikely
that more than one CV could be present in a field of a few square arcmin and
a limiting magnitude of $\le$24 mag, although a proper quantitative 
treatment would have to take into account a detailed model of the Galactic
structure. As a second point of support to our argumentation, we estimated
our detection probability of a specific post-nova system. This is based
on the limiting magnitude $m_{5\sigma}$ of our photometry, calculated as the 
faintest object that is detected in {\UBVR} and has S/N = 5 in $V$, 
corresponding to an uncertainty in the
$V$ brightness of 0.198 mag. This gives us the maximum amplitude 
$\Delta m_{5\sigma} = m_{5\sigma}-m_\mathrm{max}$ that a nova with a brightness
at maximum $m_\mathrm{max}$ can have
to still be detected in that specific photometry. Finally, the fraction of
novae with $\Delta m \le \Delta m_{5\sigma}$ translates into the
detection probability. This probability is not fixed, but depends on our
current knowledge of the $\Delta m$ distribution. As of 2016 May 15, we
count 92 pre-1986 post-novae with known $\Delta m$, with the shape of the
distribution suggesting that it is biased to bright, small-$\Delta m$ 
post-novae \citepalias{tappertetal15-2}. Our estimate furthermore 
ignores 
the possibility that a nova could have very unusual colours and thus is not 
recognized as such. 

Table \ref{prop_tab} shows that four of our objects were detected within
5 arcsec of the reported eruption and thus in its immediate vicinity. We
estimate the possibility of a different CV being present within this radius
to have a very low probability, making it unnecessary to additionally
calculate the detection probability for them. V1149 Sgr is slightly
farther off, but only V908 Oph shows a very large discrepancy, due to the
low precision of the coordinates of the nova eruption. For the photometries
of V1149 Sgr and of V908 Oph we find $m_{5\sigma}$ = 23.0 and 20.5 mag, and 
thus $\Delta m_{5\sigma}$ = 15.6 and 11.5 mag, respectively. The much lower
brightness and $\Delta m_{5\sigma}$ limit for V908 Oph is not due a lower
quality of the photometry but a result of the strong interstellar reddening
causing less than 1 per cent of the objects detected in $BV\!R$ being also
detected in $U$. The corresponding detection probabilities then calculate to
99 and 55 per cent. Additionally taking into account the small positional 
deviation to the reported coordinates, we can thus be confident that 
V1149 Sgr is indeed the post-nova. For V908 Oph, the case is less clear, and
both parameters leave room for speculation. Here, basically the only
argument supporting the identification of this system with the post-nova
of Nova Oph 1954 is the absence of other candidates in the field.

\section{Summary}
\label{sum_sec} 

\begin{enumerate}
\item We have used {\UBVR} photometry and optical spectroscopy to
recover the post-novae EL Aql, V606 Aql, V908 Oph, V1149 Sgr, V1583 Sgr
and V3964 Sgr. We furthermore report on spectroscopy of candidates for
the post-novae V1301 Aql and V1151 Sgr,
\item Historic observations of EL Aql were collected to build a complete
eruption light curve that allows it to be classified as a class P nova
\citep{stropeetal10-1}. Its eruption amplitude and colours indicate
that the previously suspected possibility that EL Aql is a recurrent nova,
is based on source confusion. Furthermore, its spectroscopic features suggest 
that the object belongs to the group of intermediate polars. 
\item Comparison of several properties of the novae indicates that the
differences in the spectroscopic appearances of EL Aql, V606 Aql, V908 Oph 
and V1149 Sgr regarding the aspect of the mass-transfer rate are likely
to be an effect of the system inclination and thus do not allow for a 
conclusive classification.
\item The shape of the H\,$\alpha$ emission line profile in V1583 Sgr indicates
that it is dominated by a non-disc emission component, likely originating in
the irradiated donor star.
\item Inspection of the emission line profiles of the individual spectra
proposes EL Aql, V606 Aql, V1583 Sgr and V3964 Sgr as good candidates for
time series observations in order to determine the orbital period.
\item We have calculated the probability that the detected objects are, in fact,
the post-novae corresponding to the reported eruptions. We have found that
the only potentially ambiguous case is that of V908 Oph, where the low
precision of the reported nova coordinates and the strong interstellar
reddening strongly reducing the number of objects detected in $U$ leave
some room for doubt. Still, in the absence of other candidates, the here
presented object should be regarded as the post-nova. 
\end{enumerate}

\section*{Acknowledgements}
This paper is based on observations with Gemini-South, program
IDs GS-2015A-Q-54 and GS-2015A-Q-75, and ESO telescopes, proposal numbers 
089.D-0505(A) and 092.D-0225(A).
We are indebted to the Gemini and ESO astronomers who performed the service 
observations. ESO furthermore provided office space and infrastructure for 
the two week stay of AE in Chile, during which large parts of the work
were discussed and prepared. Many thanks for the kind hospitality!

This research was supported by FONDECYT Regular grant 1120338 (CT and NV).
AE acknowledges support by the Spanish Plan Nacional de Astrononom\'{\i}a y 
Astrof\'{\i}sica under grant AYA2015-66211-C2-1. CT, NV and IFM acknowledge
support by the Centro de Astrof\'isica de Valpara\'iso.

We gratefully acknowledge ample use of the SIMBAD data base, 
operated at CDS, Strasbourg, France, and of NASA's Astrophysics Data System 
Bibliographic Services. {\sc iraf} is distributed by the National Optical 
Astronomy Observatories. 

The Guide Star Catalogue-II is a joint project of the Space Telescope
Science Institute and the Osservatorio Astronomico di Torino. Space
Telescope Science Institute is operated by the Association of
Universities for Research in Astronomy, for the National Aeronautics
and Space Administration under contract NAS5-26555. The participation
of the Osservatorio Astronomico di Torino is supported by the Italian
Council for Research in Astronomy. Additional support is provided by
European Southern Observatory, Space Telescope European Coordinating
Facility, the International GEMINI project and the European Space
Agency Astrophysics Division.

Gemini Observatory is operated by the Association of Universities for Research 
in Astronomy, Inc., under a cooperative agreement with the NSF on behalf of the
Gemini partnership: the National Science Foundation (United States), the 
National Research Council (Canada), CONICYT (Chile), the Australian Research 
Council (Australia), Minist\'{e}rio da Ci\^{e}ncia, Tecnologia e 
Inova\c{c}\~{a}o (Brazil) and Ministerio de Ciencia, Tecnolog\'{i}a e
Innovaci\'{o}n Productiva (Argentina).


\appendix

\section{Finding charts}

\begin{figure*}
\includegraphics[width=1.8\columnwidth]{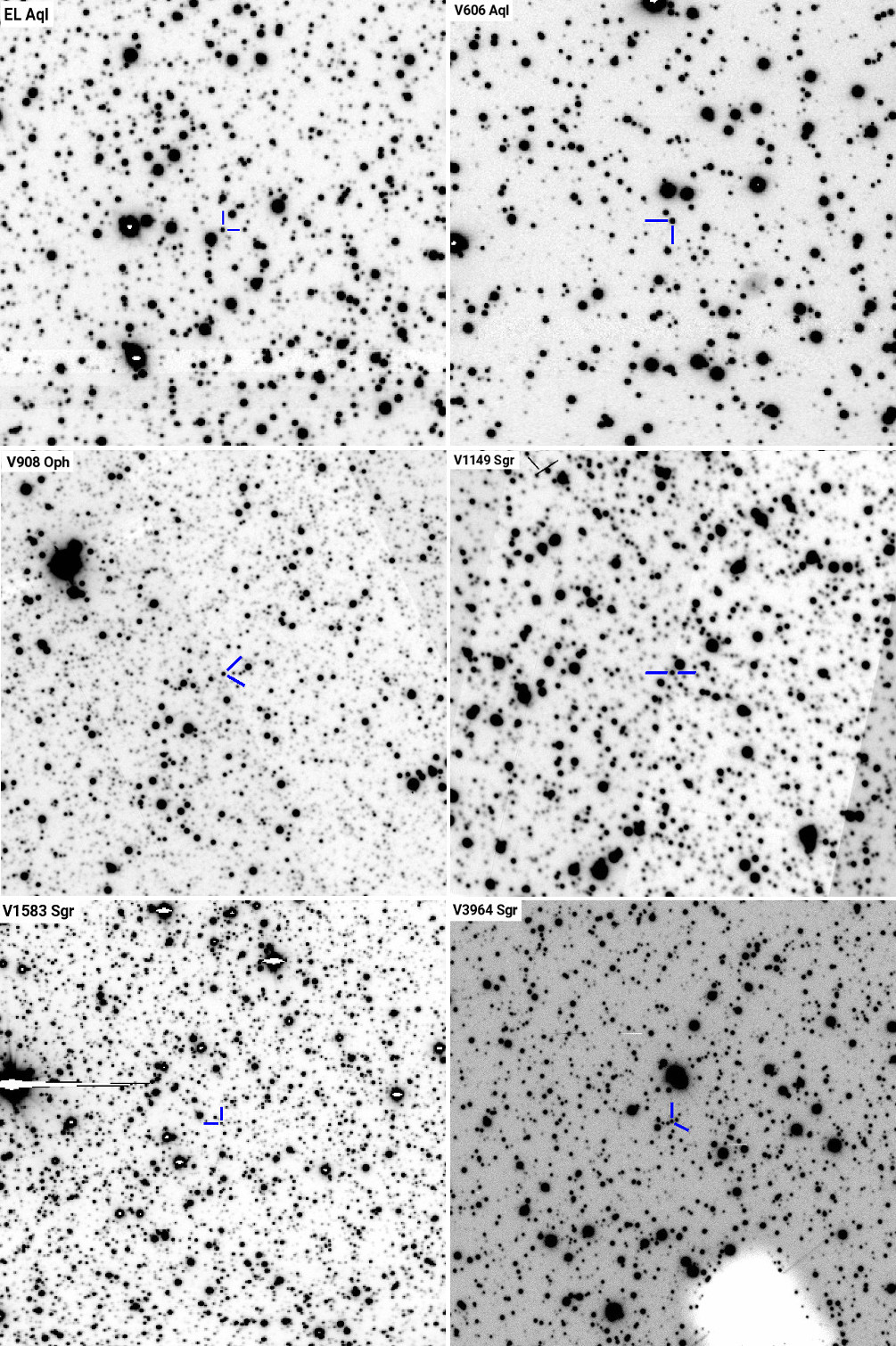}
\caption[]{
Finding charts based on images taken in $R$ or $r^\prime$ bandpasses. The size 
of a chart is 1.5 $\times$ 1.5 arcmin$^2$, and the orientation is such that 
north is up and east is to the left. The white space in the chart of V3964 Sgr 
is due to vignetting by the On-Instrument Wavefront Sensor Guiding arm.}
\label{fcs_fig}
\end{figure*}

\label{lastpage}

\end{document}